\documentstyle[12pt]{article}
\def\journal{\topmargin .3in    \oddsidemargin .5in
        \headheight 0pt \headsep 0pt
        \textwidth 5.625in 
        \textheight 8.25in 
        \marginparwidth 1.5in
        \parindent 2em
        \parskip .5ex plus .1ex         \jot = 1.5ex}
\journal

\catcode`\@=11
\def\marginnote#1{}
\catcode`\@=11
\def\section{\setcounter{equation}{0}
\@startsection {section}{1}{0pt}{-3.5ex plus -1ex minus
 -.2ex}{2.3ex plus .2ex}{\raggedright\large\bf}}
\catcode`\@=12
\newskip\humongous \humongous=0pt plus 1000pt minus 1000pt

\newif\ifdtup

\def\ident{\equiv}
\def\penclose#1{\left(#1\right)}
\def\sbenclose#1{\left[#1\right]}
\def\half{{1\over2}}

\def\e#1#2{e_{#1}^{~#2}}

\def\tlab#1#2{\tilde{L}^{#1 #2}_\infty}
\def\tp#1#2{\tilde{P}_{#1}^{~#2}}
\def\dwnup#1#2#3{{#1}_{#2}^{~#3}}
\def\alphab{\bar{\alpha}}
\newcommand{\nl}{\nonumber \\}
\def\xx{\hbox{ }^*_*}
\def\lrp{\stackrel{\leftrightarrow}{\partial}}
\def\lrbp{\stackrel{\leftrightarrow}{\bar\partial}}
\newcommand{\tr}{{\rm tr}}
\newcommand{\dif}{\partial}
\def\theequation{\thesection.\arabic{equation}}

\newtoks\stequation
\newcounter{savedequation}

\def\subequations{\refstepcounter{equation}%
  \setcounter{savedequation}{\value{equation}}%
  \stequation=\expandafter{\theequation}
  \edef\savedtheequation{\the\stequation}
  \edef\oldtheequation{\theequation}%
  \setcounter{equation}{0}%
  \def\theequation{\oldtheequation\alph{equation}}}

\def\endsubequations{\setcounter{equation}{\value{savedequation}}%
  \stequation=\expandafter{\savedtheequation}%
  \edef\theequation{\the\stequation}
  \vspace*{-12pt} \\
  }

\begin{document}
\begin{titlepage}
\begin{center}
December 1993     \hfill        UCB-PTH-93/34 \\
hep-th/9312094 \hfill     LBL-34938 \\
\hspace*{\fill} ITP-SB-93-88


{\large \bf Linearized Form of the Generic Affine-Virasoro Action}
\footnote{The work of MBH was supported in part by the Director, Office of
Energy Research, Office of High Energy and Nuclear Physics, Division of
High Energy Physics of the U.S. Department of Energy under Contract
DE-AC03-76SF00098 and in part by the National Science Foundation under
grant PHY90-21139.  The work of JdB was sponsored in part by the
National Science Foundation under grant PHY93-09888.}

\vskip .18in

J. de Boer$^1$\footnote{e-mail: DEBOER@MAX.PHYSICS.SUNYSB.EDU},
K. Clubok$^2$\footnote{e-mail: CLUBOK@PHYSICS.BERKELEY.EDU},
and M. B. Halpern$^2$\footnote{e-mail: MBHALPERN@LBL.GOV}

\vskip .18in
{\em 1. Institute for Theoretical Physics \\
State University of New York at Stony Brook\\
Stony Brook, New York 11794-3840, USA \\
\vskip .1in
2. Department of Physics \\
University of California at Berkeley \\
and \\
Theoretical Physics Group \\
Lawrence Berkeley Laboratory \\
Berkeley, CA 94720, U.S.A. \\
}

\end{center}

\vskip .15in

\begin{abstract}
Halpern and Yamron have given a Lorentz, conformal,
and Diff S$_2$-invariant world-sheet action for the
generic irrational conformal field theory, but the action is
highly non-linear.  In this paper, we
introduce auxiliary fields to find an equivalent linearized form of the action,
which shows in a very clear way
that the generic affine-Virasoro action
is a Diff S$_2$-gauged WZW model.  In
particular, the auxiliary fields transform under Diff S$_2$ as local Lie $g$
$\times$ Lie $g$ connections,
so that the linearized
affine-Virasoro action bears an intriguing resemblance to the usual
(Lie algebra) gauged WZW model.

\end{abstract}
\end{titlepage}
\setcounter{footnote}{1}

\section{Introduction}
It is now understood that affine Lie algebra [1,2] underlies
rational conformal field theory and irrational conformal field
theory (ICFT), which includes rational conformal
field theory as a small subspace.

The chiral stress tensors of ICFT are given by the general
affine-Virasoro construction [3,4]
\begin{equation}
T=L^{ab} \xx J_a J_b \xx
\end{equation}
where $J_a, a=1\ldots \dim$ g are the currents of affine g and
$L^{ab}$ is a solution of the Virasoro master equation [3,4].
See references [5,6] for reviews of recent developments in the
Virasoro master equation and the associated affine-Virasoro
Ward identities for the correlators of ICFT [7-9].

In this paper, we focus on the generic affine-Virasoro action \cite{HY},
which is a Lorentz, conformal, and Diff S$_2$-invariant world-sheet
action for the generic ICFT.  This action exhibits an elegant underlying
geometry associated with the embedding of Diff S$_2$ in (classical)
affine $G$ $\times$ $G$, but the form of the action is highly
non-linear.

The central result of this paper is an
equivalent linearized form of this action, given
in eq.(\ref{maction}),
which shows in a very clear way that the generic affine-Virasoro action
is a Diff S$_2$-gauged WZW model.
In this form, the action bears an intriguing resemblance to the ordinary
(Lie algebra) gauged WZW model [11,12], and we will see that this resemblance
is a consequence of the underlying geometry.

{}To understand the forms of the generic affine-Virasoro action, we need some
basic facts about affine-Virasoro constructions.

The original affine-Virasoro action
follows from the basic quantum
hamil\-tonian\footnote{A natural off-critical
extension of the general affine-Virasoro
construction is defined by the hamiltonian (\ref{firstham}) with
arbitrary $L^{ab}$.  These theories are scale invariant and
generically non-relativistic, generally achieving Lorentz invariance only
at the conformal points $L^{ab}$ which solve the Virasoro master
equation.} of the general theory $L$,
\begin{equation}
H_0=L_0+\bar{L}_0 = L^{ab} (\xx J_aJ_b + \bar{J}_a\bar{J}_b \xx)_0
\label{firstham}
\end{equation}
where the barred currents $\bar{J}$ and Virasoro operator
$\bar{T}=L^{ab}\xx \bar{J}_a\bar{J}_b\xx $ are right-mover copies of the
left-movers $J$ and $T$.
In general, the basic hamiltonian in (\ref{firstham}) admits a local
gauge invariance, described by the (symmetry)
algebra of the commutant of $H_0$,
and the physical (gauge-fixed) Hilbert space of the
$L$ theory may be taken as the primary states under the symmetry algebra.

As a simple example, consider the stress tensor
$T_{g/h}=L^{ab}_{g/h}\xx J_aJ_b\xx$
of the $g/h$ coset constructions
[2,13,14],
whose symmetry algebra is the affine algebra of $h$. Then the physical
Hilbert space of the non-chiral coset construction may be chosen as the
set of states
\begin{equation}
J_a{}^{m>0}|\mbox{phys}\rangle = \bar{J}_a{}^{m>0}|\mbox{phys}\rangle = 0,
\hskip 10pt a \in h
\label{ref1}
\end{equation}
which are primary under affine $h$ $\times$ $h$.

In the space of all conformal field theories,
the coset constructions are only special points of higher symmetry.
The symmetry algebra of the generic
stress tensor $T$ is the Virasoro algebra of
its commuting $K$-conjugate theory $\tilde{T}$
[2,11,12,3],
\begin{equation}
\tilde{T}=\tilde{L}^{ab}\xx J_a J_b\xx=T_g-T, \hskip 10pt
\tilde{L}^{ab}=L_g^{ab}-L^{ab}, \hskip 10pt
c(\tilde L)=c_g-c(L)
\end{equation}
where $T_g$ is the affine-Sugawara construction on g
[2,13,15,16] and $c_g$ is its central charge.
Then the physical Hilbert space of the generic theory $L$ may be chosen as the
set of states
\begin{subequations}
\begin{equation}
\tilde{L}^{m>0}|\mbox{phys}\rangle=
\bar{\tilde{L}}{}^{m>0}|\mbox{phys}\rangle=0
\end{equation}
\begin{equation}
\tilde{T}=\tilde{L}^{ab} \xx J_a J_b\xx , \hskip 15pt
\bar{\tilde{T}}=\tilde{L}^{ab}\xx \bar{J}_a \bar{J}_b \xx
\end{equation}
\label{ref2}
\end{subequations}
which are Virasoro primary under the $K$-conjugate stress tensors
$\tilde{T}$ and $\bar{\tilde{T}}$.

Transcribing (\ref{ref1}) and (\ref{ref2}) into the language of
world-sheet actions, we understand that we will obtain a spin-1 gauge
theory [11,12] for the special cases of the
coset constructions and a spin-2 gauge theory
\cite{HY}
for the generic ICFT, where the generic theory $L$ is gauged by
its $K$-conjugate theory $\tilde{L}$.

One other fact about the Virasoro
master equation is necessary to understand the generic
affine-Virasoro action.  Since an action begins as a semi-classical
description of a quantum system, the affine-Virasoro
action involves the semi-classical
(or high-level) form of the $K$-conjugate pairs of solutions of the
master equation [17,10],
\begin {subequations}
\begin{equation}
L^{ab}_\infty={1 \over 2} G^{ac}P_c{}^b, \hskip 10pt
\tilde{L}^{ab}_\infty={1 \over 2} G^{ac}\tilde{P}_c{}^b, \hskip 10pt
L^{ab}_\infty+\tilde L^{ab}_\infty=L^{ab}_{g,\infty}=\half G^{ab}
\end{equation}
\begin{equation}
c(L_\infty)=\dim P, \hskip 10pt c(\tilde L_\infty)=\dim \tilde P, \hskip 10pt
c(L_\infty)+c(\tilde L_\infty)=c(L_{g,\infty})=\dim g
\end{equation}
\begin{equation}
P^2=P, \hskip 10pt \tilde{P}^2=\tilde{P}, \hskip 10pt
P+\tilde{P}=1, \hskip 10pt P\tilde P = \tilde P P = 0.
\end{equation}
\label{mastersol}
\end{subequations}
Here, $P$ and $\tilde P$ are the high-level projectors of the $L$
and the $\tilde L$ theories respectively,
whose high-level central charges are the dimensions of their projectors.
$G^{ab}$ is the inverse of $G_{ab}$, which is the
coefficient of the central term in the general affine algebra.
The results in (\ref{mastersol}) are valid for high levels
$k_I$ of $g=\oplus_Ig_I$, but the reader may wish to
think in terms of simple g, for which
$G^{ab}=k^{-1} \eta^{ab}$ where $\eta^{ab}$ is the inverse Killing metric
of g.

Although all high-level smooth solutions of the master equation have the
form (\ref{mastersol}), it is important to recall that not all projectors
are solutions [17,10].
This is the high-level classification problem of ICFT, which
has been partially solved in terms of graph theory and generalized graph
theory [6,18].  In the graph theories, the projectors
are adjacency matrices of the graphs, each of which labels a level family of
ICFT's.

\section{Notation and the WZW Model}

We begin on the group manifold $G$ with $g(x)\in G$ in
matrix representation $T$,
\begin{equation}
[T_a,T_b] = if_{ab}{}^c, \hskip 10pt \mbox{tr} (T_aT_b)=yG_{ab}, \hskip 10pt
a,b,c=1,\ldots,\dim g
\end{equation}
and a set of coordinates $x^i(\sigma,\tau), i=1,\ldots,$ dim g.
Introduce the left-invariant and right-invariant vielbeins $e_i,\bar{e}_i$
and the antisymmetric tensor field $B_{ij}$ by
\begin{subequations}
\begin{equation}
e_i \ident -ig^{-1}\partial_i g = e_i{}^a T_a \hskip 15pt
\bar{e}_i \ident -ig\partial_i g^{-1} = \bar{e}_i{}^a T_a
\end{equation}
\begin{equation}
\partial_{[i}e_{j]}{}^a=e_i{}^b e_j{}^c f_{bc}{}^a, \hskip 15pt
\partial_{[i}\bar e_{j]}{}^a=\bar e_i{}^b \bar e_j{}^c f_{bc}{}^a
\end{equation}
\begin{equation}
\hskip 15pt i \mbox{tr}(e_i[e_j,e_k])=
     -i \mbox{tr}(\bar e_i[\bar e_j,\bar e_k])
    =\partial_i B_{jk} + \partial_j B_{ki}
                         +\partial_k B_{ij}.
\end{equation}
\end{subequations}
We also introduce inverse vielbeins $e_a{}^i$ and $\bar{e}_a{}^i$.

In this notation, the WZW action \cite{Wit} is given by,
\begin{subequations}
\begin{equation}
S_{WZW}  =  \int d\tau d\sigma ({\cal L}_{WZW} + \Gamma)
\end{equation}
\begin{equation}
{\cal L}_{WZW}  =  \frac{1}{8\pi}G_{ab} e_i{}^a e_j{}^b
   (\dot{x}^i\dot{x}^j - x'^ix'^j), \hskip 15pt
\Gamma  =  \frac{1}{4\pi y} B_{ij} \dot{x}^i x'^j
\end{equation}
\label{wzw1}
\end{subequations}
where $\Gamma$ is the WZW term.
The corresponding classical WZW hamiltonian is
\begin{subequations}
\begin{equation}
H_{WZW}=\int_0^{2\pi}d\sigma {\cal H}_{WZW}
\end{equation}
\begin{equation}
{\cal H}_{WZW} = {1 \over 2\pi} L^{ab}_{g,\infty}(J_aJ_b+\bar{J}_a\bar{J}_b),
\hskip 15pt  L^{ab}_{g,\infty} ={1 \over 2} G^{ab}
\end{equation}
\end{subequations}
where $L^{ab}_{g,\infty}$ is the high-level form of the affine-Sugawara
construction on g,
and the currents $J_a, \bar J_a, a=1,\ldots,\dim g$ are given by the
canonical representation \cite{CC},
\begin{subequations}
\begin{equation}
J_a = 2\pi e_a^{~i}\hat{p}_i + \half G_{ab} e_i^{~b}x^{\prime i},
 \hskip 15pt
\bar{J}_a = 2\pi \bar{e}_a^{~i}\hat{p}_i
        - \half G_{ab} \bar{e}_i^{~b}{x'}^i
\end{equation}
\begin{equation}
\hat{p}_i \ident p_i-{1 \over 4\pi y} B_{ij} x'^j
\label{pref}
\end{equation}
\label{currents}
\end{subequations}
which satisfies the bracket (classical) form of affine $g$ $\times$ $g$.

\section{The Generic Affine-Virasoro Action}
Assuming that the currents satisfy the algebra of (classical)
affine $g \times g$, we have the classical affine-Virasoro hamiltonian
\cite{HY},
\begin{subequations}
\begin{equation}
H=\int_0^{2\pi} d\sigma {\cal H}
\end{equation}
\begin{equation}
{\cal H}  =  {\cal H}_0(L_\infty) + v \cdot K(\tilde{L}_\infty)
\end{equation}
\begin{equation}
{\cal H}_0(L_\infty) = {1 \over 2\pi}L^{ab}_\infty(J_aJ_b+\bar{J}_a\bar{J}_b),
\hskip 15pt
v \cdot K(\tilde{L}_\infty) = {1\over2\pi}\tilde{L}^{ab}_\infty (v J_a J_b
    + \bar{v} \bar{J}_a \bar{J}_b).
\end{equation}
\label{hamil}
\end{subequations}
for the generic theory $L$, gauged by its $K$-conjugate theory $\tilde L$.
Note that all four commuting (classical) Virasoro operators
(the stress tensors),
\begin{equation}
{1\over 2\pi} L^{ab}_\infty J_a J_b, \hskip 10pt
{1\over 2\pi} L^{ab}_\infty \bar J_a \bar J_b, \hskip 10pt
{1\over 2\pi} \tilde L^{ab}_\infty J_a J_b, \hskip 10pt
{1\over 2\pi} \tilde L^{ab}_\infty \bar J_a \bar J_b
\label{stensors}
\end{equation}
are involved here, as in the quantum theory, with the semiclassical
substitution
$L\rightarrow L_\infty, \tilde{L} \rightarrow \tilde{L}_\infty$.
The first two stress tensors are used to form the classical
basic hamiltonian
\pagebreak
$H_0=\int d\sigma {\cal H}_0$ of the $L$ theory,
while the commuting
$K$-conjugate stress tensors (of the $\tilde L$ theory) play the role
of Gauss' law, and must be added, as shown, with the spin-2 gauge
field $v, \bar v$.

For passage to the action, one needs to choose a canonical representation
of the currents.  Following Ref.\cite{HY}, we choose the representation
in (\ref{currents}), but it should be emphasized that other canonical
representations can be chosen \cite{ARK} which lead to presumably
equivalent actions.

With (\ref{currents}), one obtains the hamiltonian equation of motion,
\begin{subequations}
{
\begin{equation}
\dot{x}^i  =  4\pi e_a{}^i G^{ac}M_c{}^b e_b{}^j \hat{p}_j
    + e_a{}^i G^{ab}N_b{}^c G_{cd} e_j{}^d x'^j
\end{equation}
\begin{equation}
M  \ident  1-{1-v \over 2}\tilde{P}-{1-\bar{v} \over 2}\omega\tilde{P}
          \omega^{-1}, \hskip 10pt
N  \ident  -{1-v \over 2}\tilde{P}+{1-\bar{v} \over 2}\omega\tilde{P}
          \omega^{-1}
\end{equation}
\begin{equation}
gT_a g^{-1}  = \omega_a{}^b T_b, \hskip 15pt
\omega_a{}^c G_{cd} \omega_b{}^d = G_{ab}
\end{equation}
\begin{equation}
\bar{e}_i{}^a  =  -e_i{}^b \omega_b{}^a, \hskip 15pt
\bar{e}_a{}^i=-(\omega^{-1})_a{}^b e_b{}^i
\end{equation}
}
\end{subequations}
where $\omega_a{}^b$ is the adjoint action of $g$.
Eliminating $p$ in favor of $\dot{x}$, one finds the
affine-Virasoro action of the generic theory $L$ \cite{HY},
\begin{subequations}
\begin{equation}
     S  =  \int d\tau d\sigma ({\cal L} + \Gamma)
\end{equation}
\begin{eqnarray}
     {\cal L} & = & {1\over 8\pi}\e{i}{a}G_{bc}\e{j}{c}\biggl[
        \dwnup{\sbenclose{f(Z)
          +\alpha\alphab\omega\tilde{P}\omega^{-1}f(Z)\tilde{P}}}{a}{b}
           \penclose{\dot{x}^i\dot{x}^j - x^{\prime i}x^{\prime j}}
       \nl
       & &+\alpha\dwnup{\sbenclose{f(Z)\tilde{P}}}{a}{b}
             \penclose{\dot{x}^i\dot{x}^j  + x^{\prime i}x^{\prime j}
                         +\dot{x}^{(i}x^{j)\prime}}
        \nl
      &  &+\alphab
               \dwnup{\sbenclose{\omega\tilde{P}\omega^{-1}f(Z)}}{a}{b}
             \penclose{\dot{x}^i\dot{x}^j  + x^{\prime i}x^{\prime j}
                         -\dot{x}^{(i}x^{j)\prime}}
       \nl
      & &+\dwnup{\sbenclose{1-f(Z)
          +\alpha\alphab\omega\tilde{P}\omega^{-1}f(Z)\tilde{P}}}{a}{b}
             \penclose{\dot{x}^{[i}x^{j]\prime}}\biggr]
\label{badl}
\end{eqnarray}
\label{eqafvirl}
\begin{equation}
f(Z)  \ident  [1-\alpha \alphab Z]^{-1}, \hskip 15pt
Z\ident \tilde{P}\omega\tilde{P}\omega^{-1}, \hskip 15pt
\alpha \ident {1-v \over 1+v}, \hskip 15 pt
\bar{\alpha} \ident {1-\bar{v} \over 1+ \bar{v}}.
\end{equation}
\end{subequations}
The Lorentz, diffeomorphism, and conformal symmetries of this action
are discussed in the original reference, and we confine ourselves here
to some brief remarks which will be useful below.

\noindent A.  WZW limits.
The affine-Virasoro action reduces to the WZW action
$S_{WZW}$ when we choose $L^{ab}_\infty=L^{ab}_{g,\infty}$.

For any $L^{ab}_\infty$, the action also reduces to $S_{WZW}$
in the WZW gauge,
\begin{equation}
v=\bar v=1, \hskip 10pt
\alpha=\bar\alpha=0.
\pagebreak
\end{equation}
The hamiltonian theory in this gauge must be taken with the
$K$-conjugate constraints
$\tilde L^{ab}_\infty J_aJ_b=\tilde L^{ab}_\infty\bar J_a \bar J_b =0$,
so that ${\cal H} \sim {\cal H}_0(L_\infty)$ on the constrained space.

\noindent B.  Diff S$_2(K)$ invariance.  The affine-Virasoro action is
invariant
under the Diff~S$_2(K)$ coordinate transformations,
\begin{subequations}
\begin{equation}
\delta x^i = \Lambda^a e_a{}^i + \bar\Lambda^a \bar e_a{}^i
\end{equation}
\begin{equation}
\delta g=gi\Lambda^aT_a-i\bar\Lambda^aT_ag
\label{gtrans}
\end{equation}
\begin{equation}
\delta J_a = f_{ab}{}^c \Lambda^b J_c + G_{ab}\partial_\sigma \Lambda^b,
     \hskip 10pt
\delta \bar{J}_a = f_{ab}{}^c \bar{\Lambda}^b \bar{J}_c
     -G_{ab} \partial_\sigma \bar{\Lambda}^b
\label{jtransf}
\end{equation}
\begin{equation}
\Lambda^a = 2\epsilon \tilde L^{ab}_\infty J_b, \hskip 15pt
\bar \Lambda^a = 2\bar\epsilon \tilde L^{ab}_\infty \bar J_b
\label{Lambda}
\end{equation}
\begin{equation}
\delta v  =  \dot{\epsilon} + \epsilon\lrp_\sigma v, \hskip 15pt
\delta \bar{v} = \dot{\bar{\epsilon}} + \bar{v}\lrp_\sigma \bar{\epsilon}
\label{diffv}
\end{equation}
\label{diffs2}
\end{subequations}
generated by the stress tensors of the $K$-conjugate theory.

With Ref.\cite{HY}, we remark on the embedding of Diff S$_2(K)$ in
(classical) affine $G$~$\times$~$G$.  The transformation of the
group element $g$ in eq.(\ref{gtrans}) shows that infinitesimal
Diff S$_2(K)$ transformations are particular transformations in
(classical) affine $G$~$\times$~$G$, with the current-dependent local
Lie $g$ $\times$ Lie $g$ gauge parameters $\Lambda,\bar\Lambda$ in
eq.(\ref{Lambda}).  In this sense, Diff S$_2(K)$ is embedded locally
in (classical) affine $G$~$\times$~$G$.  Moreover, the result
(\ref{jtransf}) shows that the currents $J,\bar J$ transform under
Diff S$_2(K)$ as local Lie $g$ $\times$ Lie $g$ gauge fields, or
connections, with the same gauge parameters $\Lambda,\bar\Lambda$.
The embedding of Diff S$_2$ in the affine algebra is the underlying
geometry of the generic affine-Virasoro construction, and
this geometry will continue to play
a central role in the linearized action below.

The transformation of $v,\bar{v}$ in (\ref{diffv}) allows the
identification of a second-rank tensor field,
\begin{equation}
\tilde{h}_{mn}\ident e^{-\phi}\pmatrix{ -v \bar{v} &
       {1 \over 2}(v- \bar{v}) \cr {1 \over 2}(v-\bar{v}) & 1 }, \hskip 10pt
\sqrt{-\tilde h} \tilde h^{mn} = {2 \over v+\bar v}
  \pmatrix{ -1 & \half(v-\bar v) \cr \half(v-\bar v) & v \bar v }
\label{tmetric}
\end{equation}
which was called the $K$-conjugate metric in Ref.\cite{HY}.
Note that in the WZW gauge,
\begin{equation}
\sqrt{-\tilde h}\tilde h^{mn}=\pmatrix{-1 & 0 \cr 0 & 1}
\end{equation}
so that the WZW gauge is the conformal gauge of the generic action.

{}To identify the $K$-conjugate metric
more precisely, we define the symmetric stress
tensor,
\begin{equation}
\Theta^{mn}={2\over \sqrt{-\tilde{h}}} {\delta S_{AV} \over \delta
 \tilde{h}_{mn}}
\end{equation}
which is covariantly conserved.  In the WZW gauge, we find that
\begin{subequations}
\begin{equation}
\Theta_{00}=\Theta_{11}={1\over 2\pi} \tilde{L}^{ab}_\infty
     (J_a J_b + \bar{J}_a \bar{J}_b)
\end{equation}
\begin{equation}
\Theta_{01}=\Theta_{10}={1\over 2\pi} \tilde{L}^{ab}_\infty
     (J_a J_b - \bar{J}_a \bar{J}_b)
\end{equation}
\end{subequations}
which identifies the $K$-conjugate metric $\tilde h_{mn}$
as the world-sheet metric of the $\tilde{L}$ theory.

\noindent C.  Rigid conformal invariance.
The action is invariant under a rigid (ungauged) conformal invariance,
including a rigid world-sheet Lorentz invariance, of the usual form,
\begin{subequations}
\begin{equation}
\delta \xi^\pm = -\rho^\pm(\xi^\pm), \hskip 10pt
\delta x^i = (\rho^+\partial_+ + \rho^-\partial_-) x^i
\end{equation}
\begin{equation}
\delta\alpha=(\rho^+\partial_+ + \rho^-\partial_-)\alpha
    +(\partial_-\rho^- - \partial_+\rho^+) \alpha
\end{equation}
\begin{equation}
\delta\bar\alpha=(\rho^+\partial_+ + \rho^-\partial_-)\bar\alpha
    +(\partial_+\rho^+ - \partial_-\rho^-)\bar\alpha
\end{equation}
\label{contran}
\end{subequations}
where we have defined
$\xi^\pm\ident(\tau\pm\sigma)/\sqrt{2}$.
The $\alpha, \bar\alpha$ transformations in (\ref{contran}) are
equivalent to the $v,\bar v$ transformations in eq.(4.12) of Ref.\cite{HY},
and identify $\alpha$ and $\bar\alpha$ as (-1,1) and (1,-1) conformal
tensors respectively.  With these identifications, each term in the action
density (\ref{badl}) is (1,1) on inspection.

The rigid conformal group is the conformal group of the $L$ theory,
generated by the stress tensors $L_\infty^{ab}J_a J_b/2\pi$ and
$L_\infty^{ab}\bar J_a \bar J_b/2\pi$.  The theory also has a gauged
conformal group \cite{HY}, in Diff S$_2(K)$, associated to the $K$-conjugate
stress tensors $\tilde L^{ab}_\infty J_a J_b /2\pi$ and
$\tilde L^{ab}_\infty \bar J_a \bar J_b /2\pi$.

\section{Linearized Form of the Affine-Virasoro Action}
The affine-Virasoro action in (\ref{eqafvirl}) is highly
non-linear, but we may linearize it by the introduction of auxiliary
fields.

We begin by defining two matrices,
\begin{subequations}
{\samepage
\begin{equation}
W_a^{~b} \ident \delta_a^{~b} +
   \alpha\tp{a}{b},  \hskip 15pt
(W^{-1})_a{}^b = \delta_a{}^b +{1 \over 2}(v-1)\tp{a}{b}
\end{equation}
\begin{equation}
   \dwnup{\bar{W}}{a}{b}\ident \dwnup{\delta}{a}{b}
          + \bar{\alpha}\tp{a}{b}, \hskip 15pt
(\bar{W}^{-1})_a{}^b = \delta_a{}^b + {1 \over 2}(\bar{v}-1)\tp{a}{b}.
\end{equation} }
\label{wdef}
\end{subequations}
where $\alpha, \bar\alpha$ are defined in eq.(\ref{eqafvirl}).  Using
these matrices, we add two terms to the hamiltonian and the
corresponding action of the theory,
\begin{subequations}
{
\begin{equation}
{\cal H}'  =   {\cal H} + \Delta {\cal H}, \hskip 10pt
S'  =  \int d\tau d\sigma [p_i \dot{x}^i - {\cal H}
- \Delta {\cal H}]
\end{equation}
\begin{eqnarray}
\Delta {\cal H} & = &
  -{1 \over 2\pi}\penclose{B_a - \dwnup{(W^{-1})}{a}{d}J_d}
        G^{ac}\dwnup{W}{c}{b}
     \penclose{B_b - \dwnup{(W^{-1})}{b}{d}J_d} \nl
 & & -{1 \over 2\pi}\penclose{\bar{B}_a
             - \dwnup{(\bar{W}^{-1})}{a}{d}\bar{J}_d}
        G^{ac}\dwnup{\bar{W}}{c}{b}
     \penclose{\bar{B}_b - \dwnup{(\bar{W}^{-1})}{b}{d}\bar{J}_d}.
\end{eqnarray}
\label{newform}
}
\end{subequations}
Here, $B_a$ and $\bar{B}_a,\, a=1,\ldots, \dim g$ are a set of auxiliary
fields, called the first set of
auxiliary connections for reasons discussed below.
Since the action is quadratic in $B$ and $\bar B$, the
auxiliary connections are easily integrated out, showing that the theory is
unchanged for averages of functions of $x$ and $p$.

For passage to the action, it is convenient to rearrange $H'$ as follows,
\begin{subequations}
\begin{eqnarray}
{\cal H}' &= &{\cal H}_{WZW} \nl
  &  & - {\alpha \over \pi} \tlab{a}{b}B_a B_b
   - {1\over 2\pi}(B_a - J_a)G^{ab}(B_b-J_b) \nl
  & &   - {\bar{\alpha} \over \pi} \tlab{a}{b}\bar{B}_a \bar{B}_b
       -{1\over 2\pi}
     (\bar{B}_a - \bar{J}_a)G^{ab}(\bar{B}_b-\bar{J}_b)
\label{dham}
\end{eqnarray}
\begin{equation}
\dot{x}^i=-4\pi e_a{}^i G^{ab}e_b{}^j \hat{p}_j
+2e_a{}^i G^{ab}(B_b-\omega_b{}^c \bar{B}_c).
\label{xeom}
\end{equation}
\end{subequations}
Because of the simple equation of motion in (\ref{xeom}), we obtain
our first linearized form
of the affine-Virasoro action,
\begin{subequations}
\begin{equation}
      S'  = \int d\tau d\sigma ({\cal L}' + \Gamma)
\end{equation}
\begin{eqnarray}
       {\cal L}' & = & -{1\over 8\pi} G_{ab}(\e{\tau}{a}\e{\tau}{b}
               - \e{\sigma}{a}\e{\sigma}{b})
              +{1\over\pi}G^{ab}B_a\omega_b^{~c}\bar{B}_c
        \nl
        & &+{\alpha\over \pi}\tlab{a}{b}B_a B_b
              + {1\over 2\pi} (\e{\tau}{a} - \e{\sigma}{a})B_a
        \nl
       & & +{\bar{\alpha} \over \pi} \tlab{a}{b}\bar{B}_a \bar{B}_b
                + {1\over 2\pi} (\bar\e{\tau}{a} + \bar\e{\sigma}{a})
                         \bar{B}_a
\label{aform}
\end{eqnarray}
\label{eqslamone}
\end{subequations}
where we have introduced $\e{\tau}{a}\ident\e{i}{a}\dot{x}^i,
\e{\sigma}{a}\ident\e{i}{a}x'^i$ and similarly for $\bar e$.
Note that the first term in ${\cal L}'$ and the WZW term $\Gamma$
almost comprise the action $S_{WZW}$, but the kinetic energy term has the
wrong sign.

{}To correct this sign, we introduce a second set of auxiliary connections,
\begin{equation}
A_a\ident (\omega^{-1})_a{}^b \left( B_b-{1 \over 2} G_{bc}
      (e_\tau{}^c+e_\sigma{}^c)\right), \hskip 10pt
\bar{A}_a \ident \omega_a{}^b \left( \bar{B}_b
-{1 \over 2} G_{bc} (\bar e_\tau{}^c-\bar e_\sigma{}^c)\right)
\label{curv}
\end{equation}
with $a=1,\ldots,\dim g$.
In terms of these auxiliary fields, we obtain the final form of the linearized
affine-Virasoro action,
\begin{subequations}
\begin{equation}
S'  =  S_{WZW} + \int d\tau d\sigma \Delta {\cal L}
\end{equation}
\begin{eqnarray}
  \Delta {\cal L} & = &
        {\alpha \over \pi}\tlab{a}{b} \left( \omega_a{}^c
           A_c + {1 \over 2} G_{ac}(e_\tau{}^c +e_\sigma{}^c) \right)
      \left( \omega_b{}^d
           A_d + {1 \over 2} G_{bd}(e_\tau{}^d +e_\sigma{}^d) \right) \nl
      & + &{\bar{\alpha}\over \pi} \tlab{a}{b} \left( (\omega^{-1})_a{}^c
         \bar{A}_c +{1 \over 2} G_{ac}(\bar e_\tau{}^c-\bar e_\sigma{}^c)
         \right) \left( (\omega^{-1})_b{}^d
         \bar{A}_d +{1 \over 2} G_{bd}(\bar e_\tau{}^d-\bar e_\sigma{}^d)
         \right)  \nl
     & + &{1\over\pi}\bar A_a G^{ab} \omega_b^{~c}A_c
\label{fform}
\end{eqnarray}
\label{gform}
\end{subequations}
where $S_{WZW}$ is the WZW action in (\ref{wzw1}).  In this form, we may
easily verify that $S'$ reduces to $S_{WZW}$ in the WZW gauge, since the
auxiliary fields essentially decouple.

For the reader's convenience, we also give this action in terms of the
group variable $g$,
\begin{subequations}
{\samepage
\begin{equation}
S'=S_{WZW}+\int d^2z \Delta {\cal L}
\end{equation}
\begin{equation}
S_{WZW}=-{1 \over 2\pi y}\int d^2z \, \tr(g^{-1}\partial g g^{-1}
                   \bar\partial g)
       -{1 \over 12\pi y}\int_M \tr(g^{-1}dg)^3
\end{equation}
\begin{eqnarray}
\Delta{\cal L}& = &
     -{\alpha \over \pi y^2} \tilde L^{ab}_\infty
     \tr \left(T_ag^{-1}Dg\right)
     \tr \left(T_bg^{-1}Dg\right) \nl
\nopagebreak
 & & -{\bar\alpha \over \pi y^2} \tilde L^{ab}_\infty
     \tr \left(T_ag\bar Dg^{-1}\right)
     \tr \left(T_bg\bar Dg^{-1}\right) \nl
\nopagebreak
 & & +{1\over \pi y} \tr(g\bar A g^{-1} A)
\label{mform}
\end{eqnarray}
\begin{equation}
A\ident A_aG^{ab}T_b, \hskip 10pt \bar A\ident \bar A_aG^{ab}T_b, \hskip 10pt
\partial\ident\half(\partial_\tau+\partial_\sigma), \hskip 10pt
\bar\partial\ident\half(\partial_\tau-\partial_\sigma)
\end{equation}
\begin{equation}
D\ident\partial+iA,\hskip 15pt \bar D\ident\bar\partial+i\bar A
\end{equation}
\label{maction}
}
\end{subequations}
where we have defined the covariant derivatives
$D,\bar D$ and $d^2z=d\tau d\sigma$.

In this form, the theory is clearly seen as a Diff S$_2$-gauged WZW model,
which bears an intriguing resemblance to the form of the usual
(Lie algebra) gauged WZW model [11,12].  As we shall see in the
following section, this resemblance is due to the Diff S$_2(K)$
transformations of the auxiliary connections.

It is straightforward to check that the linearized action
is equivalent to
the non-linear form of the affine-Virasoro action.  Using the
inversion formula
\begin{subequations}
\begin{equation}
     C_A{}^B  = \pmatrix{\alpha\tp{a}{b} & \dwnup{\delta}{a}{b} \cr
                      \dwnup{\delta}{a}{b} &
                   \alphab(\omega\tilde P\omega^{-1})_a{}^b},
\hskip 20pt
A,B=1,\ldots,2 \dim g
\end{equation}
\begin{equation}
   (C^{-1})_A{}^B  =
         \pmatrix{-\alphab
              \left(\omega\tilde{P}\omega^{-1} f(Z)\right)_a{}^b
                           &\left(1+\alpha\alphab
                  \omega\tilde{P}\omega^{-1} f(Z)\tilde{P}\right)_a{}^b
                           \cr   f(Z)_a{}^b
              & -\alpha \left(f(Z) \tilde{P}\right)_a{}^b}
\end{equation}
\end{subequations}
to integrate out the auxiliary fields in (\ref{eqslamone})
or (\ref{gform}), we obtain exactly
(\ref{eqafvirl}) in the form
\begin{subequations}
\begin{equation}
     S = \int d\tau d\sigma ({\cal L}+\Gamma)
\end{equation}
\begin{equation}
     {\cal L}  =
           -{1\over 8\pi}G_{ab}
                   \penclose{\e{\tau}{a}\e{\tau}{b}
                           - \e{\sigma}{a}\e{\sigma}{b}}
         -{1\over 8\pi}
               E^A  (C^{-1})_A{}^B E_B
\end{equation}
\begin{equation}
E^A= \pmatrix{(\e{\tau}{a}-\e{\sigma}{a}),
           & (\bar\e{\tau}{a}+\bar\e{\sigma}{a})}, \hskip 15pt
E_B= \pmatrix{G_{bc}(\e{\tau}{c}-\e{\sigma}{c})
           \cr G_{bc}(\bar\e{\tau}{c}+\bar\e{\sigma}{c})}.
\end{equation}
\end{subequations}

\section{Invariances of the Linearized Action}

Because it is simpler, we discuss first the rigid conformal invariance,
which is the conformal invariance of the $L$ theory.  We find that the
actions (\ref{eqslamone}) and (\ref{gform}) or (\ref{maction})
are invariant under the
transformations (\ref{contran}), supplemented by the conformal
transformations of the auxiliary connections,
\begin{subequations}
\begin{equation}
\delta B_a=(\rho^+\partial_+ + \rho^-\partial_-)B_a + \partial_+\rho^+B_a,
\hskip 10pt
\delta \bar B_a=(\rho^+\partial_+ + \rho^-\partial_-)\bar B_a
   + \partial_-\rho^-B_a
\end{equation}
\begin{equation}
\delta A=(\rho^+\partial_+ + \rho^-\partial_-)A + \partial_+\rho^+A,
\hskip 10pt
\delta \bar A=(\rho^+\partial_+ + \rho^-\partial_-)\bar A
   + \partial_-\rho^-A
\end{equation}
\end{subequations}
which identify  $A$ and $B$ as (1,0) tensors and $\bar A$ and $\bar B$ as
(0,1) tensors.  With these identifications, each term in (\ref{aform}),
(\ref{fform}), and (\ref{mform}) is a (1,1) tensor on inspection.

For Diff S$_2(K)$, we find that the action (\ref{eqslamone})
is invariant under

\begin{subequations}
\begin{equation}
\delta x^i = \lambda^a e_a{}^i + \bar\lambda^a \bar e_a{}^i
\end{equation}
\begin{equation}
\delta \alpha   =  -\bar\partial\xi+\xi\lrp\alpha, \hskip 15pt
\delta \bar\alpha = -\partial\bar\xi + \bar\xi\lrbp\bar\alpha
\label{alpha}
\end{equation}
\begin{equation}
\delta B_a = f_{ab}{}^c \lambda^b B_c + G_{ab}\partial\lambda^b, \hskip 10 pt
\delta \bar B_a = f_{ab}{}^c \bar\lambda^b \bar B_c
      + G_{ab}\bar\partial\bar\lambda^b
\label{atrans}
\end{equation}
\begin{equation}
\lambda^a  \ident  2\xi\tilde L^{ab}_\infty B_b, \hskip 10 pt
\bar\lambda^a \ident 2\bar\xi\tilde L^{ab}_\infty \bar B_b
\label{lambda}
\end{equation}
\label{newtrans}
\end{subequations}
where $\xi,\bar\xi$ are the diffeomorphism parameters.
The $\alpha, \bar\alpha$ diffeomorphisms in (\ref{alpha}) are
equivalent to the $v,\bar v$ diffeomorphisms in (\ref{diffv}) under the
parametric redefinition
\begin{equation}
 \xi = (1+\alpha) \epsilon, \hskip 15pt
    \bar\xi= (1+\bar\alpha) \bar\epsilon
\label{idents}
\end{equation}
where $\epsilon, \bar\epsilon$ are the diffeomorphism parameters
in (\ref{diffs2}).

The action (\ref{gform}) or (\ref{maction})
is also invariant under Diff S$_2(K)$. For the vielbein form of the final
action in (\ref{gform}),
the transformation
\begin{equation}
\delta A_a=f_{ab}{}^c\bar\lambda^b A_c+G_{ab}\partial\bar\lambda^b,
\hskip 10pt
\delta \bar A_a=f_{ab}{}^c\lambda^b\bar A_c+G_{ab}\bar\partial\lambda^b
\label{btrans}
\end{equation}
replaces that of $B$ in (\ref{atrans}).  For the $g$ form of the
final action in (\ref{maction}), the
transformations
\begin{subequations}
\begin{equation}
\delta g=gi\lambda-i\bar\lambda g
\end{equation}
\begin{equation}
\delta A=\partial\bar\lambda+i[A,\bar\lambda], \hskip 10pt
\delta \bar A=\bar\partial\lambda+i[\bar A,\lambda]
\end{equation}
\begin{equation}
\lambda\ident\lambda^aT_a,\hskip 15pt \bar\lambda\ident\bar\lambda^aT_a
\end{equation}
\label{mtrans}
\end{subequations}
replace those for $x$ and $B$ in (\ref{newtrans}).
In (\ref{btrans}) and (\ref{mtrans}), it is understood
that the gauge parameters
$\lambda^a(B(A))$ and $\bar\lambda^a(\bar B(\bar A))$ are expressed
in terms of the connections
$A$ and $\bar A$, using eqs.(\ref{lambda}) and the inverse of (\ref{curv}).

We remark in particular that the auxiliary connections
in (\ref{atrans}), (\ref{btrans}),
and (\ref{mtrans}) transform under Diff S$_2(K)$
as local
Lie $g$ $\times$ Lie $g$ connections,
with field-dependent local Lie $g$ $\times$ Lie $g$ parameters
$\lambda, \bar\lambda$, restricted as shown in (\ref{lambda}).
As emphasized above, and in Ref.\cite{HY}, this
is possible because Diff S$_2(K)$ is locally embedded in (classical)
affine $G \times G$.
The Diff S$_2(K)$ transformations of the auxiliary connections are also
responsible for the intriguing
resemblance of the linearized action (\ref{gform}) or (\ref{maction})
to the form of the usual (Lie algebra) gauged WZW model [11,12].

Because verification of the
diffeomorphism invariance of the linearized actions involves
considerable algebra, we give some useful intermediate results
for the $g$ form of the final action in (\ref{maction}).  We need
\begin{equation}
\delta(g^{-1}Dg)=\partial(i\lambda)+[g^{-1}Dg,i\lambda], \hskip 10pt
\delta(g\bar Dg^{-1})=\bar\partial(i\bar\lambda)
    +[g\bar Dg^{-1},i\bar\lambda]
\end{equation}
and, using the explicit form of $\tilde L^{ab}_\infty$ in (\ref{mastersol}),
we obtain the term-by-term results,
\begin{subequations}
\begin{equation}
\delta S_{WZW}  = -\frac{i}{\pi y} \int d^2 z \,
\tr(g^{-1} \dif g \bar{\dif}\lambda
- \bar{\dif} g  g^{-1} \dif \bar{\lambda} )
\end{equation}
\begin{equation}
\delta  \left(
\frac{1}{\pi y} \int d^2 z \, \tr
(g\bar{A} g^{-1} A) \right)  =
\frac{1}{\pi y} \int d^2 z \, \tr (g^{-1} A g \bar{\dif}\lambda +
g \bar{A} g^{-1} \dif \bar{\lambda} )
\end{equation}
\begin{equation}
\delta
\left( -\frac{1}{\pi y^2} \int d^2 z \,
\alpha \tilde{L}_{\infty}^{ab}
\tr (T_a g^{-1} D g ) \tr (T_b g^{-1} D g )  \right)  =
\frac{i}{\pi y} \int d^2 z \, \tr (g^{-1} D g \bar{\dif}\lambda )
\end{equation}
\begin{equation}
\delta
\left( -\frac{1}{\pi y^2} \int d^2 z \,
\alpha \tilde{L}_{\infty}^{ab}
\tr (T_a g\bar{D} g^{-1} ) \tr (T_b g \bar{D} g^{-1} )
\right)   =
\frac{i}{\pi y} \int d^2 z \, \tr (
g \bar{D}  g^{-1} \dif \bar{\lambda} )
\end{equation}
\end{subequations}
whose sum $\delta S'$ is zero on inspection.

It should also be emphasized that the non-linear and linear forms of the
affine-Virasoro action are conformal and Diff S$_2(K)$-invariant because
$L^{ab}_\infty$ and $\tilde L^{ab}_\infty$ are solutions of the high-level
Virasoro master equation.  For example, temporarily assuming arbitrary
$\tilde L^{ab}_\infty$, we obtain the extra terms
in the $\bar\lambda$ variation of $S'$,
\begin{eqnarray}
\delta S'&=&{-2\over\pi y^2}\int d^2z \, \bar\alpha
      \tr(g^{-1}\bar DgT_a)
      (\tilde L^{ab}_\infty-2\tilde L^{ac}_\infty G_{cd} \tilde L^{db}_\infty)
      \partial\left[\bar\xi\tr(g^{-1}\bar DgT_b)\right] \nl
    &=& 0
\end{eqnarray}
which vanish for the explicit solutions $\tilde L^{ab}_\infty$ in
(\ref{mastersol}).

\section{The Doubly-Gauged Action}

So far, we have been discussing the world-sheet action of the $L$
theory, which is gauged by its commuting $K$-conjugate theory
$\tilde L$, and we have seen that the spin-2 gauge field
$\tilde h_{mn}$ of this theory is the world-sheet metric of
the $\tilde L$ theory.  Halpern and Yamron have also indicated how to
incorporate the world-sheet metric of the $L$ theory,
\begin{equation}
h_{mn}\ident e^{-\chi}\pmatrix{ -u \bar{u} &
       {1 \over 2}(u- \bar{u}) \cr {1 \over 2}(u-\bar{u}) & 1 }, \hskip 10pt
\sqrt{-h} h^{mn} = {2 \over u+\bar u}
  \pmatrix{ -1 & \half(u-\bar u) \cr \half(u-\bar u) & u \bar u }
\end{equation}
which results in the {\em doubly-gauged} affine-Virasoro action,
with two world-sheet metrics, $h_{mn}$ and
$\tilde h_{mn}$ in (\ref{tmetric}).  We will sketch these
results quickly since the steps are analogous to those above for the
singly-gauged theory.

One begins with the doubly-gauged hamiltonian \cite{HY},
\begin{subequations}
\begin{equation}
H_D=\int d\tau d\sigma {\cal H}_D
\end{equation}
\begin{equation}
{\cal H}_D = {1\over 2\pi} \left[
  ( u L_\infty + v \tilde L_\infty )^{ab} J_a J_b +
      ( \bar{u} L_\infty + \bar{v} \tilde L_\infty )^{ab}
         \bar{J}_a \bar{J}_b \right ]
  \ident u\cdot T + v\cdot K
\end{equation}
\end{subequations}
which reduces to the hamiltonian (\ref{hamil}) of the $L$ theory when
$u=\bar u=1$.  Defining the matrices,
\begin{subequations}
\begin{equation}
W_a^{~b} \ident \delta_a^{~b} +
   \alpha\tp{a}{b} + \beta P_a{}^b,  \hskip 15pt
(W^{-1})_a{}^b = \delta_a{}^b +{1 \over 2}(v-1)\tp{a}{b}
      +\half(u-1)P_a{}^b
\end{equation}
\begin{equation}
   \dwnup{\bar{W}}{a}{b}\ident \dwnup{\delta}{a}{b}
          + \bar{\alpha}\tp{a}{b} + \bar\beta P_a{}^b, \hskip 15pt
(\bar{W}^{-1})_a{}^b = \delta_a{}^b + {1 \over 2}(\bar{v}-1)\tp{a}{b}
      +\half(\bar u-1)P_a{}^b
\end{equation}
\begin{equation}
\alpha={1-v\over 1+v}, \hskip 15pt
\bar\alpha={1-\bar v \over 1+ \bar v}, \hskip 15pt
\beta\ident{1-u\over 1+u}, \hskip 15pt
\bar\beta\ident{1-\bar u \over 1+ \bar u}
\end{equation}
\end{subequations}
we follow the prescription in (\ref{newform}) to
construct the modified hamiltonian density,
\begin{eqnarray}
{\cal H}_D' &= &{\cal H}_{WZW} \nl
  &  & - {1 \over \pi}  (\alpha \tilde L_\infty + \beta L_\infty)^{ab} B_a B_b
 - {1\over 2\pi}(B_a - J_a)G^{ab}(B_b-J_b) \nl
 & & -{1 \over \pi} (\bar\alpha L_\infty +\bar\beta \tilde L_\infty)^{ab}
                   \bar{B}_a \bar{B}_b
   -{1\over 2\pi} (\bar{B}_a - \bar{J}_a)G^{ab}(\bar{B}_b-\bar{J}_b)
\end{eqnarray}
where $B$ and $\bar B$ are the first set of auxiliary connections.
The linearized forms of the doubly-gauged action are then obtained
by the simple substitution
\begin{equation}
\alpha\tilde L^{ab}_\infty \rightarrow \alpha\tilde L^{ab}_\infty
                                      +\beta L^{ab}_\infty, \hskip 15pt
\bar\alpha\tilde L^{ab}_\infty \rightarrow \bar\alpha\tilde L^{ab}_\infty
                                      +\bar\beta L^{ab}_\infty, \hskip 15pt
\end{equation}
in the singly-gauged actions (\ref{eqslamone}), (\ref{gform}), or
(\ref{maction}).

Similarly, we we may integrate out the auxiliary connections of the
linearized forms to obtain the non-linear form of the doubly-gauged action,
\begin{subequations}
\begin{equation}
     S_D = \int d\tau d\sigma ({\cal L}_D+\Gamma)
\end{equation}
\begin{equation}
     {\cal L}_D  =
           -{1\over 8\pi}G_{ab}
                   \penclose{\e{\tau}{a}\e{\tau}{b}
                           - \e{\sigma}{a}\e{\sigma}{b}}
         -{1\over 8\pi}
               E^A  (C^{-1})_A{}^B E_B
\end{equation}
\begin{equation}
E^A= \pmatrix{(\e{\tau}{a}-\e{\sigma}{a}),
           & (\bar\e{\tau}{a}+\bar\e{\sigma}{a})}, \hskip 15pt
E_B= \pmatrix{G_{bc}(\e{\tau}{c}-\e{\sigma}{c})
           \cr G_{bc}(\bar\e{\tau}{c}+\bar\e{\sigma}{c})}
\end{equation}
\begin{equation}
  C= \pmatrix{\alpha\tilde{P} + \beta P & 1 \cr
         1 & \omega(\alphab\tilde{P}+\bar{\beta} P)\omega^{-1}}
\end{equation}
\begin{equation}
  C^{-1}     = \pmatrix{-\omega(\alphab\tilde{P}+\bar{\beta} P)\omega^{-1} f
         &1+\omega(\alphab\tilde{P} + \bar\beta P)
                  \omega^{-1} f(\alpha\tilde{P}+\beta P)\cr
                  f &  -f(\alpha\tilde{P}+ \beta P)}
\end{equation}
\begin{equation}
f \ident \sbenclose{1-(\alpha\tilde{P}+\beta P) \omega
         (\alphab\tilde{P}+\bar{\beta} P) \omega^{-1}}^{-1}
\end{equation}
\end{subequations}
It is not difficult to check that this result
agrees with the original form (\ref{eqafvirl}) of the
affine-Virasoro action
when $u=\bar u=1$ so that $\beta=\bar\beta=0$.

The doubly-gauged actions are invariant under
Diff S$_2(T)$ $\times$ Diff S$_2(K)$, which is generated
by all four stress tensors in (\ref{stensors}).  The
Diff S$_2(T)$ transformations of the new Diff S$_2(T)$ gauge fields are
\begin{subequations}
\begin{equation}
\delta u=\dot\kappa+\kappa\lrp_\sigma u, \hskip 15pt
\delta \bar u  = \dot{\bar\kappa}+\bar u \lrp_\sigma \bar\kappa
\end{equation}
\begin{equation}
\delta\beta=-\bar\partial\zeta+\zeta\lrp\beta,\hskip 15pt
\delta\bar\beta=-\partial\bar\zeta+\bar\zeta\lrbp\bar\beta
\end{equation}
\begin{equation}
\zeta=(1+\beta)\kappa, \hskip 15pt \bar\zeta=(1+\bar\beta)\bar\kappa
\end{equation}
\end{subequations}
which is a copy of eqs.(\ref{diffv}), (\ref{alpha}), and (\ref{idents}),
in terms of
the new Diff S$_2(T)$ gauge parameters $\kappa,\bar\kappa$ and
$\zeta,\bar\zeta$.  The transformations of $x$, $g$, and the connections
$A,\bar A$ or $B, \bar B$ may be obtained
by the substitution
\begin{equation}
\lambda^a \rightarrow \lambda^a= 2(\xi \tilde L^{ab}_\infty
             + \zeta L^{ab}_\infty) B_b, \hskip 15pt
\bar\lambda^a \rightarrow \bar\lambda^a= 2(\bar\xi \tilde L^{ab}_\infty
             + \bar\zeta L^{ab}_\infty) \bar B_b
\label{newlambda}
\end{equation}
for the local Lie $g$ $\times$ Lie $g$ gauge parameters $\lambda,\bar\lambda$
in eqs.(\ref{newtrans}), (\ref{btrans}), and (\ref{mtrans}).
The forms in (\ref{newlambda}) show that Diff S$_2(T)$ $\times$
Diff S$_2(K)$ is also embedded in (classical) affine $G$~$\times$~$G$.

\section{Conclusions}
Some time ago, Halpern and Yamron found the generic affine-Virasoro action,
which is a Lorentz, conformal, and Diff S$_2$-invariant world-sheet action
for the generic irrational conformal field theory.  This action exhibits
an elegant underlying geometry associated with the embedding of Diff S$_2$
in (classical) affine $G$~$\times$~$G$, but the form of the action is
highly non-linear.

In this paper, we introduced auxiliary fields $A_a, \bar A_a,
a=1,\ldots,\dim g$ to find the linearized form of the affine-Virasoro
action, given in
eq.(\ref{maction}).  The embedding of Diff S$_2$ in (classical)
affine $G$ $\times$ $G$ plays a particularly transparent role in the
linearized action, which is clearly seen as a Diff S$_2$-gauged
WZW model.
In particular,
the auxiliary fields $A,\bar A$
transform under Diff S$_2$ as local Lie $g$ $\times$
Lie $g$ connections, which accounts for the
intriguing resemblance of the linearized
action to the usual (Lie algebra) gauged
WZW model.

A next step is to integrate out the spin-2 gauge fields of the non-linear or
the linearized action, following the usual semiclassical development
in two-dimensional gravity.
In this case, we are gauging a $K$-conjugate ``matter'' system with
semiclassical central charge $c(\tilde L_\infty)=\dim \tilde P$,
so we expect the
semiclassical action to have the form \cite{DI}
\begin{subequations}
\begin{equation}
S= S_{WZW}+S_G+S_{FF}
\end{equation}
\begin{equation}
c(L_{g,\infty})= \dim g, \hskip 10pt
c_G=-26, \hskip 10pt
c_{FF}=1-12Q^2=26-c(\tilde L_\infty)
\end{equation}
\end{subequations}
in the WZW gauge $\alpha=\bar\alpha=0$.
Here, $S_G$ is the action of the
diffeomorphism ghosts and $S_{FF}$ is a Feigen-Fuchs action with
its background charge adjusted so that the gauged subsystem ($K$-conjugate
matter + ghosts + Feigen-Fuchs)
has zero semiclassical
central charge.  Then we may verify that
the total semiclassical central charge
of the theory is the semiclassical central charge of the $L$ theory,
\begin{equation}
c({\rm total})=c(L_{g,\infty})+c_G+c_{FF}=c(L_{g,\infty})-c(\tilde L_\infty)
=c(L_\infty)=\dim P
\end{equation}
as it should be.

A full quantum version of this picture may be realized with standard
BRST operators of the form,
\begin{equation}
Q=\oint{dz\over2\pi i} c\left(\tilde T+T_{FF}+\half T_G\right), \hskip 10pt
c_{FF}=26-c(\tilde L)
\end{equation}
where $\tilde T=T(\tilde L)$ is the quantum stress tensor of the $\tilde L$
theory, with quantum central charge $c(\tilde L)$.
Up to discrete states, these BRST
operators reproduce the physical state conditions (\ref{ref2}) for the
generic theory $L$.

After completion of this work, we received a paper by Tseytlin \cite{ARK}
which studies an ungauged but presumably related action with
chiral group variables and dim $g$ auxiliary fields.  The hamiltonian of this
action is the ungauged basic affine-Virasoro
hamiltonian $H_0$ in (\ref{hamil}), with a different
canonical representation of the currents, so the gauged form of this action
should be equivalent to the results presented here.

\section*{Acknowledgments}

We thank M. W. Craig, who participated in the early states of
this work.

The work of MBH was supported in part by the Director, Office of
Energy Research, Office of High Energy and Nuclear Physics, Division of
High Energy Physics of the U.S. Department of Energy under Contract
DE-AC03-76SF00098 and in part by the National Science Foundation under
grant PHY90-21139.  The work of JdB was supported in part by the
National Science Foundation under grant PHY93-09888.

\end{document}